\documentclass[aps,prab,twocolumn,superscriptaddress,floatfix]{revtex4-1}

\usepackage{amssymb}
\usepackage{lipsum}
\usepackage{amssymb}
\usepackage{bbold}
\usepackage{chngcntr}
\usepackage{amssymb}
\usepackage{graphicx}
\usepackage{graphics}
\usepackage{amsmath}\allowdisplaybreaks
\usepackage{siunitx}
\usepackage{placeins}
\usepackage{hyperref}
\usepackage[dvipsnames]{xcolor}
\usepackage{wasysym}
\usepackage{soul}
\usepackage{float}
\usepackage{array}
\usepackage{tabulary}
\usepackage{caption}
\usepackage{fancyhdr}
\usepackage{wrapfig}
\usepackage{mathtools}
\usepackage[version=4]{mhchem}

\usepackage[normalem]{ulem}
\usepackage{graphicx}
\usepackage{color,soul}
\usepackage{subfigure}
\usepackage{dcolumn}
\usepackage{bm}
\usepackage{hyperref}
\usepackage{url}
\usepackage{multirow}
\hypersetup{
    colorlinks=true,
    linkcolor=blue,
    filecolor=blue,      
    urlcolor=blue,
    citecolor=blue,
}

\usepackage[makeroom]{cancel}

\begin{document}

\title{Heating of \ce{Cs2Te} Photocathode via Field Emission and RF Pulsed Heating: Implication Toward Breakdown} 

    \author{\firstname{Ryo} \surname{Shinohara$^*$}}
    \affiliation{Department of Electrical and Computer Engineering, Michigan State University, MI 48824, USA}
    \affiliation{Department of Physics and Astronomy, Michigan State University, East Lansing, MI 48824, USA}
    \affiliation{Theoretical Division, Los Alamos National Laboratory, Los Alamos, NM 87545, USA}
    \email{shinoha4@msu.edu}
    \author{\firstname{Soumendu} \surname{Bagchi}}
    \affiliation{Center for Nanophase Materials Sciences, Oak Ridge National Laboratory, Oak Ridge, TN 37831}
    \author{\firstname{Evgenya} \surname{Simakov}}
    \affiliation{Accelerator Operations and Technology Division, Los Alamos National Laboratory, Los Alamos, NM 87545, USA}
    \author{\firstname{Danny} \surname{Perez}}
    \affiliation{Theoretical Division, Los Alamos National Laboratory, Los Alamos, NM 87545, USA}
        \author{\firstname{Sergey V.} \surname{Baryshev$^\dagger$}}
	\affiliation{Department of Electrical and Computer Engineering, Michigan State University, MI 48824, USA}
	\affiliation{Department of Chemical Engineering and Material Science, Michigan State University, MI 48824, USA}
 \email{serbar@msu.edu}

\begin{abstract}
The occurrence of radio-frequency (rf) breakdown limits operational electromagnetic gradients in accelerator structures. Experimental evidence often suggests that breakdown events are associated with temperature and dark current spikes on the surface of radiofrequency (RF) devices. In the past decade, there has been increased interest in unveiling the mechanism behind breakdown initiation in metal copper and copper alloys; however,  effort regarding breakdown phenomenon in photocathode-relevant semiconductors have been more limited.
In this work, we explore field-emission-assisted heating via Nottingham and Joule processes, as a possible candidate for breakdown initiation. 
For this, field emission from intrinsic Cs$_2$Te ultra-thin film coated on a copper substrate was modeled within the Stratton–Baskin–Lvov–Fursey formalism, describing the processes and effects in the bulk and on the surface of a photocathode exposed to high radio-frequency electromagnetic fields.
It is shown that field emission characteristic deviates significantly from the classical Fowler-Nordheim (FN) theory, whereby predicting that dark current is orders of magnitude lower than one expected by FN law. Conventional pulsed heating was also found to impose negligible heating to the photocathode. 
Both conclusions suggest that Cs$_2$Te photocathode coated on a metal substrate would be insensitive to catastrophic thermal-material runaway breakdown, unlike what is observed for metal surfaces. 
Finally, a few unconventional breakdown candidate scenarios are identified and discussed including thermo-elastic deformation and avalanche breakdown.
\end{abstract}

\maketitle

\section{Introduction}\label{S:intro}
Semiconductors like \ce{Cs2Te} are gaining interest as electron emitters in radio-frequency (RF) photoinjectors for high-power accelerators and scientific instruments, such as time-resolved electron microscopes and free electron lasers. Compared to metal cathodes, \ce{Cs2Te} offers advantages like low intrinsic emittance and high quantum efficiency \cite{instruments8010019, Alexander:2023squ}.
In photoinjectors, higher operating fields are desired to enhance beam brightness, which can lead to new research frontiers by enhancing the signal to noise ratio or temporal resolution.
However, RF breakdown acts as a primary factor limiting the maximal operating fields that can be sustained. Breakdowns can significantly affect the structure and morphology of inner walls that face high electric fields (including photocathode surface), eventually leading to either gradual or instant performance degradation \cite{simakov2018advances}.

Experimental observations indicate that breakdowns are typically accompanied by dark current and temperature spikes. A hypothesis then can be formed that elevated/spiking temperature is a direct consequence of the dark current spike through explosive field emission at nano-asperities \cite{RFPulsedHeating}. This triggers a feedback loop leading to a catastrophic thermal runaway on the surface; which can culminate in a fully developed breakdown event, ending with the formation of a vacuum arc and power quenching. Additionally, the frequency at which breakdowns occur is very sensitive to the RF magnetic field component mediated via so-called pulsed heating effects \cite{pulsed-heating}. While the underlying breakdown nucleation mechanisms are extremely complex, considerable progress has been made in unveiling various aspects of the breakdown initiation process in metals over the past decade \cite{2018Kyritsakis, Descoeudres2009, Eimrel2015, Hebrew2019, atomistic-coupling2022,electrodiffusion2023,Shinohara2024}; to date, corresponding investigations in semiconductors have been comparatively lacking.


Recently, high gradient C-band ($f_{rf}$$\sim$5.7 GHz) accelerators were identified as promising systems for industry, medicine, national security, and basic sciences. These systems strike a balance between being relatively compact (as compared to L- or S-band) while still being able to transport high charge/high current particle beams with minimal wakefield/breakup instabilities (as compared to X-band). Future technological requirements point to the following operating ranges: electric gradient $E$ up to 300 MV/m, magnetic field $H$ up to 500 kA/m, pulse length $\tau_p$ between 0.4 and 1 $\mu$s and repetition rate $f_r$ up to 200 Hz \cite{Schneider2022, Alexander:2023squ}. This pushes the material accelerator cavities are made of into unprecedented conditions, requiring careful assessment in terms of breakdown susceptibility. 

In this work, we investigate the relative contributions of both field emission and pulsed heating due to high magnetic field on \ce{Cs2Te}-on-Cu photocathode heating to delineate the regimes that could potentially lead to a runaway breakdown process. Our findings suggest that field/dark emission has an overall limited contribution to the heating of the \ce{Cs2Te} surface due to the charge depletion characteristic of any intrinsic or low doped semiconductor. 
In contrast, photocathode heating caused by the surface magnetic field was much stronger in the range of the studied electric field/magnetic field/repetition rate parameter space, causing temperature elevation high enough to be studied experimentally $in$ $situ$. Hence, the presented results provide an avenue to refine and validate the commonly assumed mechanisms of the physics of breakdown.


\section{Computational approach and methods}


\begin{figure}
	\includegraphics[width=8.6cm]{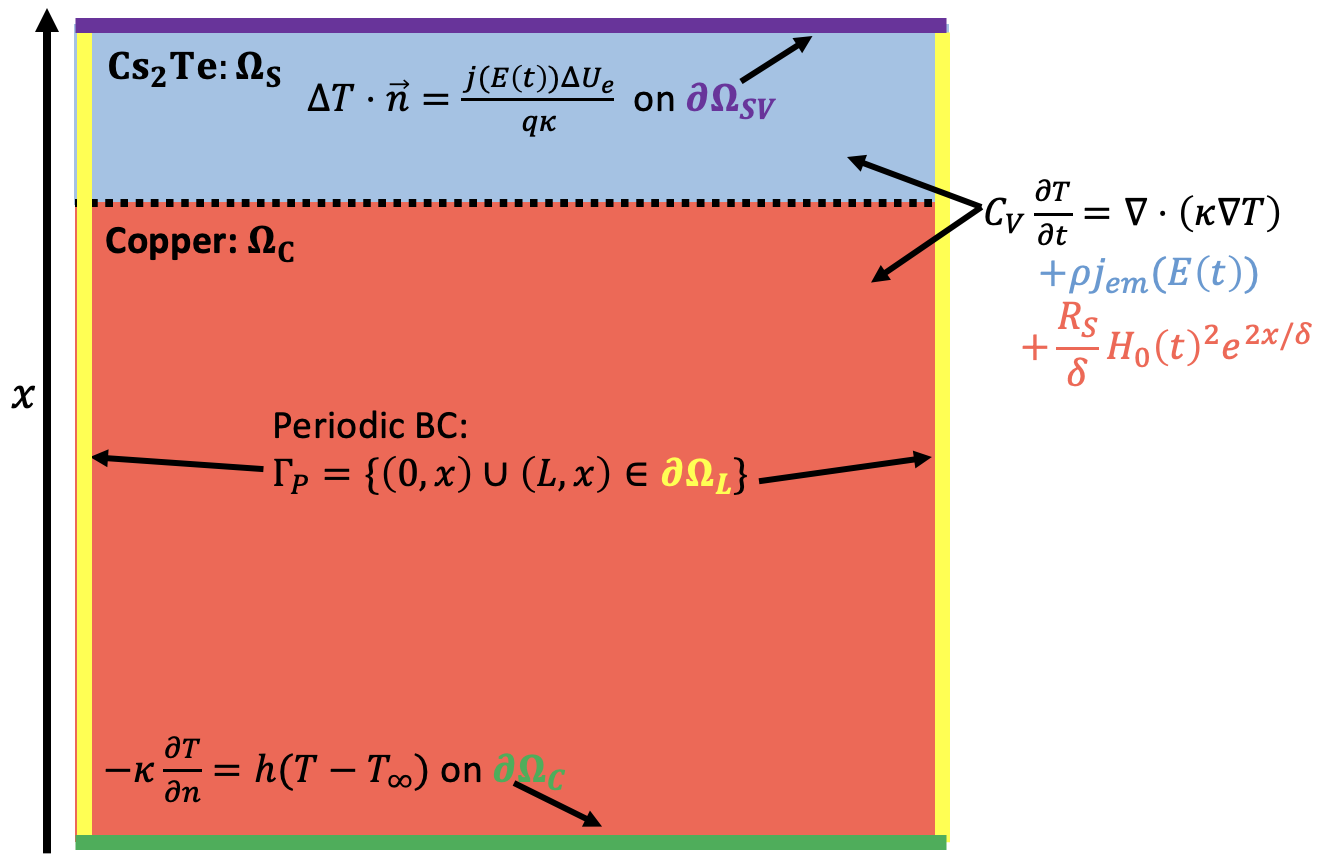}
	\caption{Schematic of 2D domain with copper subdomain $\Omega_C$ in red and \ce{Cs2Te} subdomain $\Omega_S$ in blue. The corresponding partial differential equations and boundary conditions are shown.}
	\label{fig:sim-setup}
\end{figure}

\subsection{Simulation Setup}
A system comprised of an ultrathin 50 nm \ce{Cs2Te} film grown on a Cu semi-infinite substrate within the described operating envelope was analytically modeled with respect to two key mechanisms: field-emission-induced heating and RF pulsed heating.

In order to analyze the contribution of field emission and pulsed heating to the temperature evolution in the \ce{Cs2Te}-Cu system, we consider a 2D domain with a Cu subdomain $\Omega_{C}$ (shown in red) and a \ce{Cs2Te} subdomain $\Omega_{S}$ (shown in blue), as shown in Fig. \ref{fig:sim-setup}. Transient electromagnetic (EM) fields were considered, meaning that the field amplitude during a RF pulse is described by the fill time $\tau$ of the cavity characterizing how quickly the cavity reaches the maximal field and then empties. The field amplitude evolution during the RF pulse was given by the following set of expressions:
\begin{equation} \label{eq:EM(t)}
\begin{aligned}
H_0(t)=
    \begin{cases}
        H_0(1-e^{-\frac{t}{\tau}}), &  t < t_{p}\\
        H_0(1-e^{-\frac{t_{p}}{\tau}})e^{-\frac{t-t_{p}}{\tau}}, & t > t_{p}
    \end{cases} \\
E_0(t)=
    \begin{cases}
        E_0(1-e^{-\frac{t}{\tau}}), &  t < t_{p}\\
        E_0(1-e^{-\frac{t_{p}}{\tau}})e^{-\frac{t-t_{p}}{\tau}}, & t > t_{p}
    \end{cases}
\end{aligned}
\end{equation}
where $t_p$ is the pulse length. For $t<t_p$, the field exponentially saturates to maximum field of $E_0$ or $H_0$, and for $t>t_p$ the field will display exponential decay. It is important to note that $H_0(t)$ and $E_0(t)$ are effective RMS values over RF cycles during a pulse. This simplification is valid because the time-scale of heat diffusion (roughly $L^2 \kappa/C\approx1\mu$s) is significantly larger than an a RF cycle (1/$f_{rf}$ in ns).

The temperature in the system can be solved within the heat diffusion problem following:
 \begin{equation} \label{eq:heat_diffusion}
    C_V \frac{\partial T}{\partial t} = \nabla\cdot(\kappa\nabla T) + f 
\end{equation}
where $C_V$ is the volumetric heat capacity [$\textrm{J}\cdot \textrm{K}^{-1}\cdot \textrm{m}^{-3}$], $\kappa$ is the heat conductivity, and $f$ is the volumetric heating power density, which acts as a source term for the heat equation.

\subsection{Field Emission}\label{field_emission}
\begin{figure}
	\includegraphics[width=8.6cm]{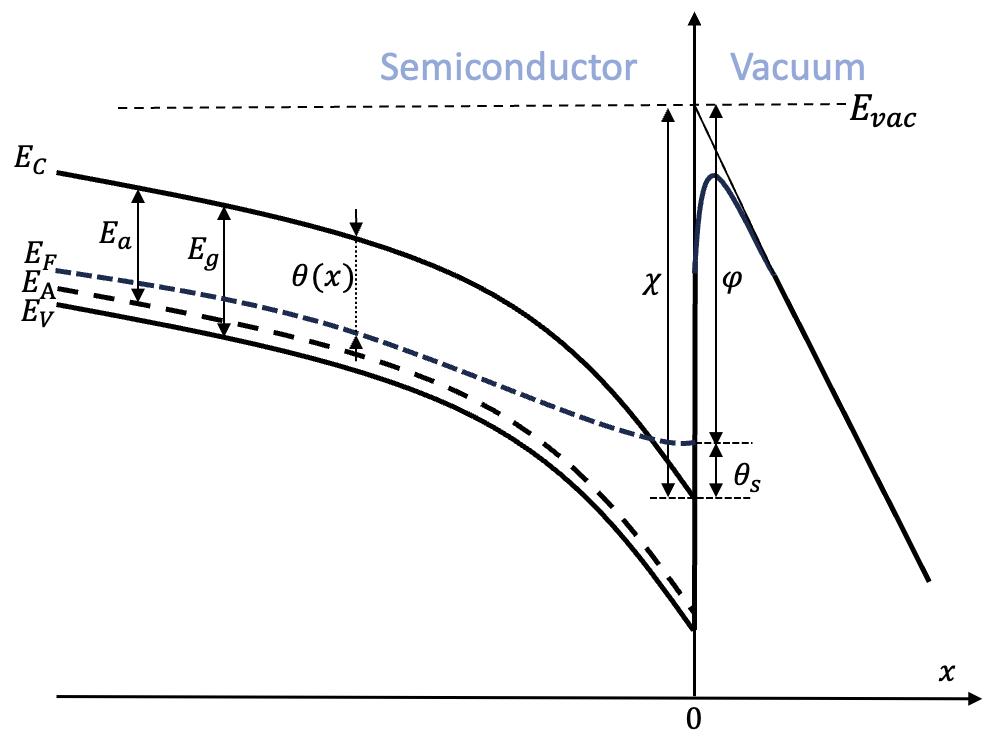}
	\caption{Energy  diagrams of semiconductor emitter in a strong electric field for a $p$-type semiconductor such as \ce{Cs2Te}. See text for details.}
	\label{fig:band_diagram}
\end{figure}

Experiments shows that the field emission current characteristic of a semiconductor significantly deviates  from the classical FN theory \cite{CarbonFilm2002,DiamondFilm2011,CarbonNanotube1998,Serbun_silicon2012}, which is commonly used to predict field emission from metals. Another theory that was developed well beyond the classical FN theory, known as 
Stratton–Baskin–Lvov–Fursey (SBLF) formalism \cite{Oksana_UNCD, Baskin1971}, proved to be indispensable when modeling field emission from semiconductors. 
Unlike in metals, when a semiconductor is placed in a strong enough electric field, that field penetrates to certain depth and depletes that region from charge. The charge can be moved across the system with a latency associated with the transit time across the depleted depth. As shown in Fig. \ref{fig:band_diagram}, in 1D energy-coordinate representation,  field penetration may cause the conduction band minimum (CBM) to bend bellow the Fermi level, allowing for an inversion layer to form and creating a well where electrons can accumulate near the  surface of the semiconductor, thereby effectively "metallizing" the surface. The tunneling probability from that well then defines the initial phase of field emission current, often appearing as if the current-field characteristic follows the conventional FN law. However, as the tunneling probability increases with the increase of the surface field, the electron well quickly depletes and new electrons have to be resupplied to sustain a high field-emission current. Current saturation is expected when field emission becomes resupply-limited due to $(i)$ limited amount of charge, $(ii)$ limited drift velocity and $(iii)$ distance/time to drift.

\subsubsection{Stratton–Baskin–Lvov–Fursey formalism}
To describe this picture self-consistently, a series equations must be solved together with special boundary conditions. A schematic energy diagram of the band bending region specific for a $p$-type semiconductor (such as \ce{Cs2Te}) is shown in Fig. \ref{fig:band_diagram}. Here, $E_C$ and $E_V$ are the conduction band minimum (CBM) and valence band maximum (VBM) respectively. $E_F$ is the Fermi level and $E_A$ is the acceptor level. $E_g$ is the band gap, $E_a$ is the energy of the acceptor level with respect to the CBM, $\chi$ is the electron affinity, $\varphi$ is the work function, and $\theta(x)$ defines the energy of the CBM with respect to the Fermi level.

The band-bending of the near surface region (the space-charge region) of the semiconductor can be described by solving the reformulated Poisson equation as a function of a dimensionless variable $y=\frac{\theta (x)}{k_B T}=\frac{E_F(x)-E_C(x)}{k_B T}$\cite{Oksana_UNCD, Baskin1971}:
\begin{subequations}
\begin{align} 
    F = -dV/dx \label{eq:poisson}\\
    \frac{dF}{dy} = -\frac{k_BT/q}{k\epsilon_0}\left[\frac{-q(n-p+N^-+N^+)}{F-j/[q(n\mu_e+p\mu_h)]}\right] \label{eq:poisson_solve}
\end{align}
\end{subequations}
Here, $F$ is the local electric field and $n,p,N^-,N^+$ are the charge densities of the electron, hole, acceptors, and donors respectively. When the densities are defined as the function of dimensionless variable $y$, the equations become:

\begin{equation}
\begin{aligned} 
   n(y) = N_C F_{1/2}(y) \\
   p(y) = N_V F_{1/2}(\frac{E_g}{k_B T}-y) \\
   N^-(y) = \frac{N_a}{1+2\exp (-E_a/k_BT-y)}\\
   N^+(y) = \frac{N_d}{1+2\exp (E_d/k_BT+y)} \label{eq:charge_density}
\end{aligned}
\end{equation}
Here, $F_{1/2}(\eta)=2/\sqrt{\pi}\int_0^\infty\xi^{1/2}[1+e^{\xi-\eta}]^{-1}d\xi$ is the Fermi-Dirac integral of order 1/2.

Solving Eq. \ref{eq:poisson_solve} yields the relationship between the electric field and $y(x)$ for a given constant current $j$ flowing through the material. The boundary condition $y_b(x_b)$ can then be determined from the value of the field $F_b$ at the conductor-semiconductor boundary where $F_b$ can be found by solving Ohm's equation in the bulk as $F_b=j/[q(n(y_b)\mu_e+p(y_b)\mu_p)]^{-1}$.

In order to obtain the field emission current, one needs to simultaneously solve the Poisson equation and the Stratton equation (which gives the emission current density from the conduction band) \cite{Baskin1971}. When the emission current $j_{em}$ is held fixed, we can retrieve the band-bending on the surface $y_s$ as a function of the surface field $F_s$. With negative surface levels, $y(F_s)<0$, the Stratton equation can be written as a simplified Nordheim equation \cite{Baskin1971, Oksana_UNCD, simplified_nordheim}
\begin{equation}
\begin{aligned}\label{eq:stratton_neg}
    y_s^-(F_s)=\ln\left(\frac{j_{em}}{C_1}\right)+C_2 \frac{\chi^{3/2}}{F_s}\zeta(F_s) \\
    \zeta(F_s)=1-Y^2+\frac{1}{3}Y^2 \ln(Y)\\
    Y=\frac{1}{\varphi}\sqrt{\frac{q F_s}{4 \pi \epsilon_0}\left( \frac{\kappa-1}{\kappa+1} \right)},
\end{aligned}
\end{equation}
while with positive surface levels, $y(F_s)>0$, the Stratton equation is written as
\begin{align}\label{eq:stratton_pos}
    j_{em}(F_s,y_s)=&\frac{q^3F_s^2}{8\pi h \varphi}\exp\left( -C_2\frac{\varphi^{3/2}}{F_s}\zeta(F_s) \right) \times \notag \\
    &\left[ 1-\exp\left(-C_3 \frac{\varphi^{1/2}}{F_s} y_s\right) \right. \notag\\
     &\left. -C_3\frac{\varphi^{1/2}}{F_s} y_s  \exp\left(-C_3\frac{\varphi^{1/2}}{F_s}y_s\right)\right]
\end{align}
where $C_1=4 \pi m_0 q^3 (k_B T)^2/h^3$ A/m$^2$, $C_2=$ $8\pi \sqrt{2 m_0 q}/(3h)$ V eV$^{-3/2}$ m$^{-1}$, and $C_3=4\pi\sqrt{2 M_0} k_B T/$ $(qh)$ V eV$^{-1/2}$ m$^{-1}$. The positive Stratton equation is solved numerically for the band bending $y^+_s(F_s)$ with fixed $j_{em}$.

\begin{figure}
	\includegraphics[width=8.6cm]{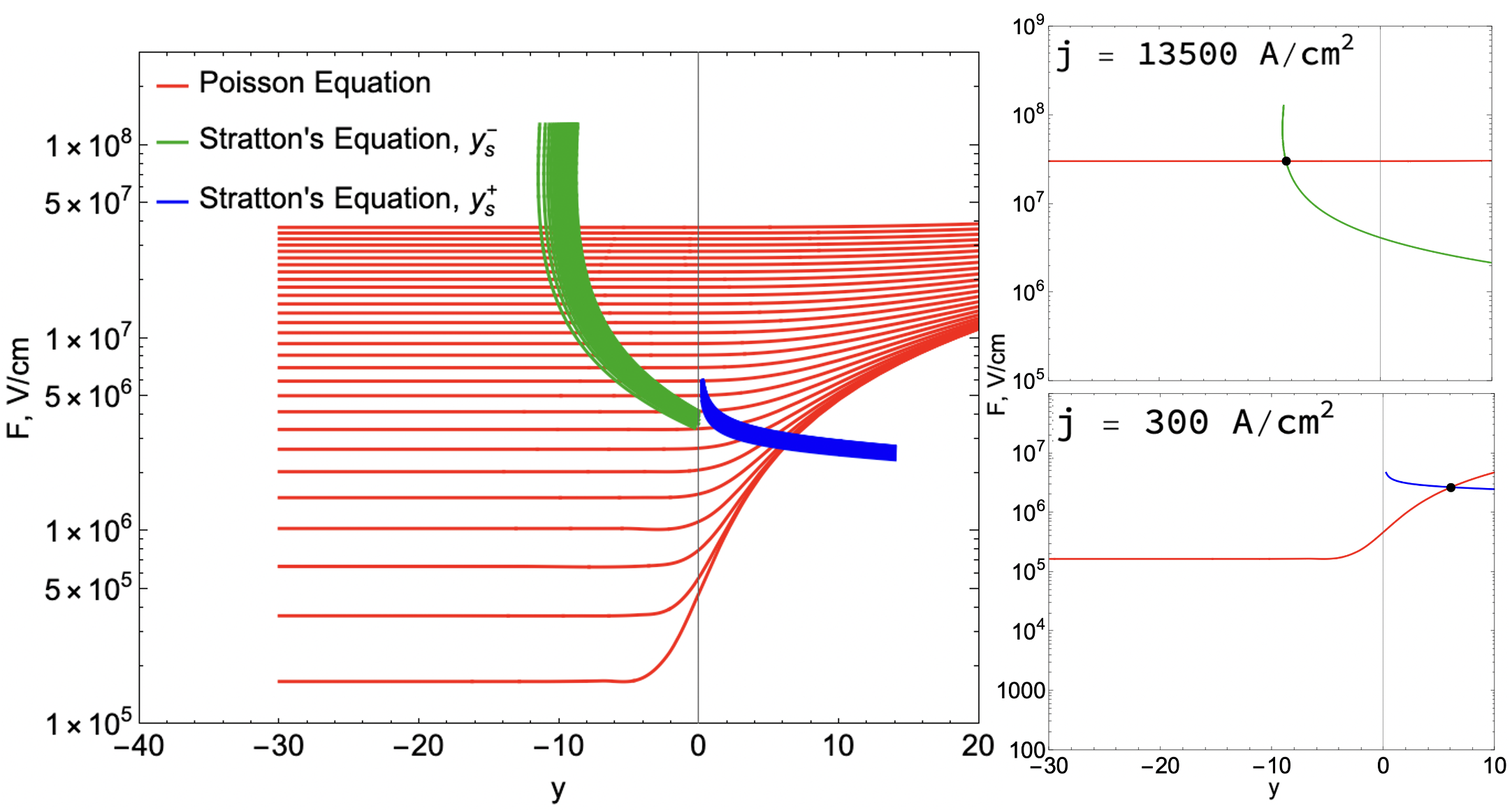}
	\caption{Solutions to the Poisson Eq. \ref{eq:poisson_solve} shown in red and to the positive and negative Stratton's Eqs. \ref{eq:stratton_neg} and \ref{eq:stratton_pos} shown in blue and green respectively. An example of intersection for each Stratton equation is plotted on the right panels.} \label{fig:SBLF_eg}
\end{figure}

The $j-E$ curve is obtained from solving multiple sets of coupled Poisson and Stratton equations, where each set has different value for the fixed $j$ and $j_{em}$. The two equations give the electric field as a function of dimensionless variable $y$ where the intersection of the curves provide the current-density dependencies on the electric field as plotted in Fig. \ref{fig:SBLF_eg}. Each intersection becomes a data point along the $j-E$ curve. 
The SBLF current profile was then mapped to the temporal electric field variation during a pulse as defined in Eq.~\ref{eq:EM(t)}.

The current profile retrieved from the SBLF theory is finally used to calculate:

\noindent 1) The Joule heating generated in the \ce{Cs2Te} layer ($\Omega_S$), where the volumetric heat source in Eq.\ref{eq:heat_diffusion} is $f(x,t)=\rho_e j^2_{em}(E(t))$ with $\rho_e$ being the electrical resistivity, and

\noindent 2) The Nottingham heating generated in the \ce{Cs2Te} layer ($\Omega_S$), where the volumetric heat source is $f(x,t)=\Delta U_e j_{em}(E(t))$ with $\Delta U_e$ being the electron energy change (release or absorption) at the Vacuum-\ce{Cs2Te} interface (see below) \cite{Nottingham}.

\subsection{RF Pulsed Heating}\label{sec:pulsed_heating}
Pulsed heating arises from the ohmic losses of surface eddy currents induced by the time-dependent RF magnetic field. Here, the pulsed heating is considered to act solely in the Cu subdomain because 
thin-film \ce{Cs2Te} will be invisible to the tangential magnetic field. Indeed, in the GHz range, the magnitude of power dissipation in an insulator or semiconductor is measured through the loss tangent 
calculated as \cite{Baker2012}
$\tan(\delta_d)=\frac{\varepsilon''}{\varepsilon'}\approx\frac{\sigma}{\omega \varepsilon_0 \varepsilon_r}$ with $\varepsilon''$ and $\varepsilon'$ the material complex permittivity $\varepsilon_c=\varepsilon-i\frac{\sigma}{\omega}=\varepsilon'-i\varepsilon''$, and $\delta_d$  the loss angle. Non-intentionally doped wideband gap semiconductor like \ce{Cs2Te} should naturally have  high resistivity ($\sim$$100~\Omega$ m) and high relative permittivity (10.17) \cite{materials_project} , leading to a very low estimate the loss tangent on the order of $10^{-2}$  under operating frequencies in GHz. From this, power dissipation and in turn heat generation in the semiconductor layer should be negligibly small. Indeed, once the loss tangent is taken account, a rudimentary calculation with $\Delta T = P_{ave}\cdot R_{thermal}$ shows that the relative temperature rise in the Cs$_2$Te layer to be on the order of 1 K. Direct dissipation in the semiconductor was thus neglected in the following.


In the metal substrate, the power density dissipated is of the form \cite{Jackson1998}
 \begin{equation}\label{eq:P(x,t)}
     P(x,t)=\frac{R_s}{\delta}|H_0(t)|^2 e^{-\frac{2x}{\delta}}
 \end{equation}
 where $R_s=\frac{\rho_e}{\delta}$ is the surface resistance and $\delta$ is the skin depth. 
To estimate the corresponding temperature rise, the diffusion Eq.\ref{eq:heat_diffusion} was solved  with $f(x,t)=P(x,t)$ as a volumetric heat source acting only in the Cu subdomain. 

\subsection{Spatio-temporal Heating Calculations}
\subsubsection{Variational Formulation of Heat Diffusion Equation}
The heat diffusion equation was solved on a 2D domain using the finite-element method (FEM). This required turning Eq.~\ref{eq:heat_diffusion} into its variational formulation and discretizing it under a timestep  $\Delta t$. 
Let $t^n$ denote the time at \textit{n}th time step with $T^n$ being the temperature field at time $t^n$. Using the backward Euler method, the time derivative can be approximated as: $\frac{\partial T}{\partial t}\bigr|_{t=t^n}=\frac{T^{n+1}-T^n}{\Delta t}$. This equation can be written in the weak formulation as $a(T,v) = L_{n+1}(v)$ by multiplying the PDE with a test function $v \in \hat{V}(\Omega)$, and integrating the second order derivative by parts in domain $\Omega$. If we consider both Neumann boundary condition 
on $\partial \Omega_{SV}$ and Robin boundary condition 
on $\partial \Omega_C$, the following equation can be obtained

\begin{equation}
\begin{aligned}\label{eq:variational}
    a(T,v) &=\int_\Omega (C_V T^{n+1} v + \kappa \Delta t \nabla T^{n+1}\cdot\nabla v )dx+\\
    &\Delta t \int_{\partial \Omega_R} (h T^{n+1} v)ds \\
    L_{n+1}(v) &= \int_\Omega C_V (T^n + \Delta t f^{n+1} ) v dx + \\ 
    &\Delta t \int_{\partial \Omega_N}(q^{n+1} v)ds + \Delta t \int_{\partial \Omega_R}(h T_\infty v)ds
\end{aligned}
\end{equation}
This resulting governing equation was solved utilizing the open source PDE solver FEniCS \cite{fenics}. FEniCS includes a domain-specific language called Unified Form Language (UFL) where the variational forms are defined. The governing equations were automatically compiled and executed in underlying C++ based computational kernel DOLFIN \cite{dolfin}.


\subsubsection{Field Emission and Pulsed Heating Simulation}

In order to define the temperature distribution $T(t,x,y)$ in a 2D domain, the variational formulation of the diffusion Eq. \ref{eq:variational} was simultaneously solved in the two domains $\Omega_{S}$ and $\Omega_C$, as defined in Fig. \ref{fig:sim-setup}. In this case, field emission induced heating in the \ce{Cs2Te} domain self-consistently with pulsed heating in the Cu domain. Here, field-emission current generates heat through two mechanism: Joule and Nottingham heating. Joule heating is taken into account by defining the deposited heat from the field emission current density as $f=\rho_e \cdot j^2$. The Nottingham effect \cite{Nottingham1936} causes heating at the semiconductor-vacuum boundary $\partial \Omega_{SV}$ resulting from the energy difference between the electron emitted from the semiconductor-vacuum interface and the electron replacing the emitted electron arriving from the metal-semiconductor interface, labeled as $\Delta U_e$. 
Near room temperatures, emitted electron would typically have lower energy compared to the replaced electron resulting in heating to take at the boundary. The Nottingham effect  on $\partial \Omega_{SV}$ can be accounted for as a Neumann boundary condition \cite{Jensen2008, Nottingham1936}
\begin{equation}
    \int_{\partial \Omega_{SV}} (\Delta T \cdot \vec{n})v ds = \int_{\partial \Omega_{SV}} \bigg(\frac{j \Delta U_e}{q \kappa}\bigg)v ds. \label{eq:nottingham}
\end{equation}

As explained above in Sec. \ref{sec:pulsed_heating}, the volumetric heat source on the Cu subdomain $\Omega_C$.  is derived from Eq. \ref{eq:P(x,t)}. To simulate the heat transfer from the copper to the surrounding, a Robin boundary condition is implemented on the bottom of the copper domain to simulate convective cooling as
\begin{equation}
\begin{aligned}
    &-\kappa \frac{\partial T}{\partial n} = h(T-T_\infty)\\
    &- \int_{\partial \Omega_C}[h(T-T_\infty)]v ds = \int_{\partial \Omega_C}hT_\infty vds - \int_{\partial \Omega_C}hT vds,
\end{aligned}
\end{equation}
where $h$ is the convection heat transfer coefficient and $T_\infty$ is the temperature of the surrounding. In this work, $h = 1.2\times 10^4$ W m$^{- 2}$ K$^{- 1}$ was used, corresponding to a water cooling channel often used in RF accelerators \cite{RFPulsedHeating}.

\section{Results and discussion}

\subsection{Field Emission Currents}
\begin{figure}
	\includegraphics[width=8.6cm]{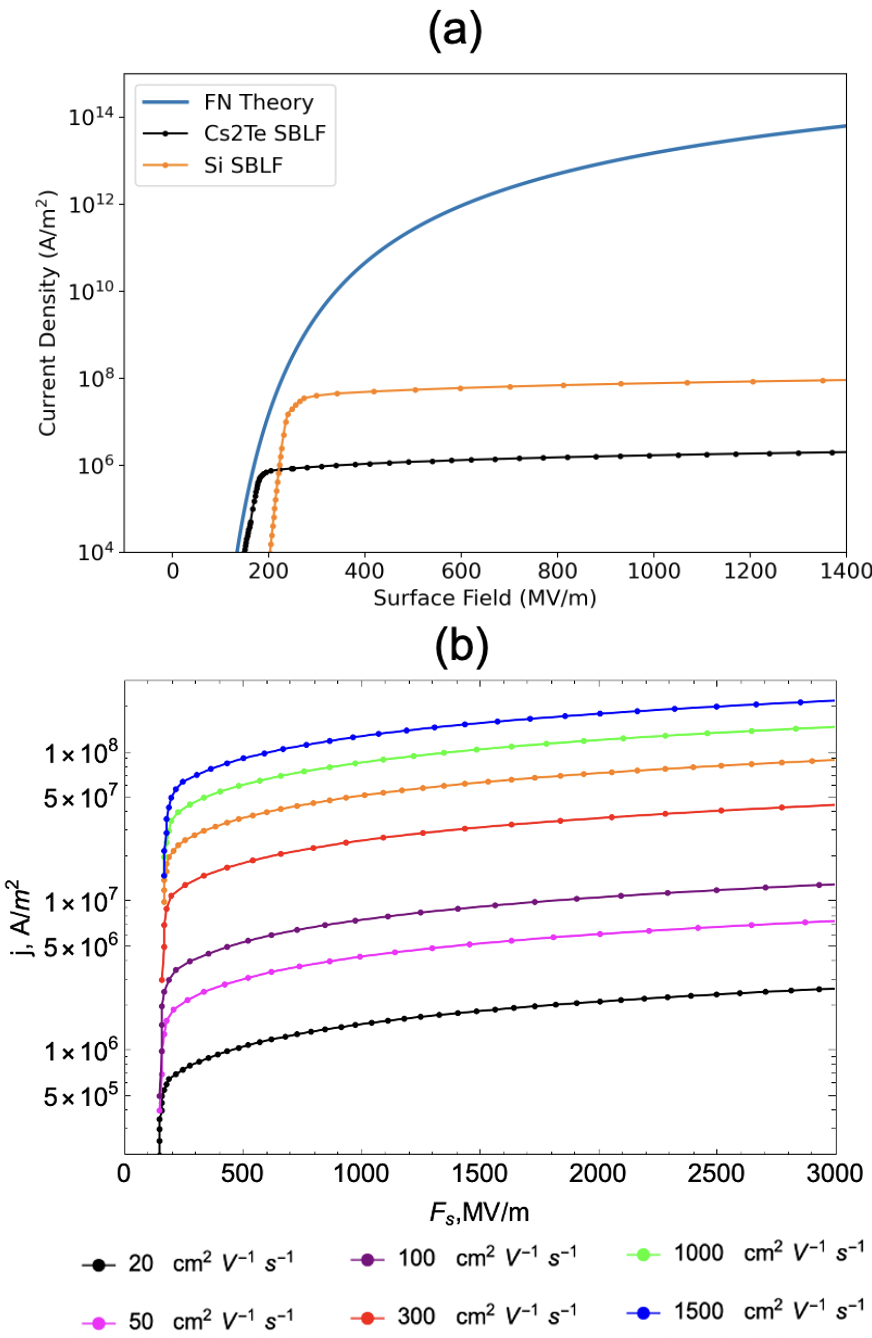}
	\caption{(a) Calculated field emission current density for \ce{Cs2Te} and Si. FN prediction plotted in blue.
    ; (b) Parametrized field emission current density for known range of electron mobility for semiconductors (\ce{Cs2Te}: 20, \ce{Si}: 1500 cm$^2$V$^{-1}$s$^{-1}$).}
	\label{fig:fe_result}
\end{figure}

The dependence of the current density  on the surface field, calculated using the FN and SBLF formalism are shown in Fig. \ref{fig:fe_result}(a). The results highlight a striking difference. A rapid current density saturation at fields above 200 MV/m was observed within SBLF formalism.  In contrast, predictions from the FN formalism (for a work function of 3 eV),  predicted current densities 6 to 8 orders of magnitude larger for the same range of the surface field. It is also noted that there is a narrow range of fields ($j=10^4 - 10^6$ A/m$^2$) where the FN and SBLF predictions are similar. This is the manifestation of the metal-like inversion layer discussed above. While the field emission current profile of \ce{Cs2Te} is currently not known in detail, we can look to other similar materials to validate our results. Chubenko $et$ $al.$ utilized a modified SBLF formalism to study field emission from $n$-type ultra-nanocrystaline diamond films, which showed good agreement with experiments \cite{Oksana_UNCD}. Field emission from $p$-type Si tip emitters was well characterized, e.g., by Serbun $et$ $al.$ \cite{Serbun_silicon2012}. 
Additional $j-E$ curves in Fig. \ref{fig:fe_result}(a) are theoretical SBLF calculations for Si emitters as described in Ref.\cite{Serbun_silicon2012}. Here, the saturation current density calculated by the SBLF theory is a much better predictor than the conventional FN theory. With rough estimate of dividing the current by the total surface area of the Si-tips, we arrive at approximate experimental current density saturation of $\sim$1e7 A/m$^2$ where classical FN theory will massively over-estimate the current density by multiple order of magnitude. 
Similar current density saturation levels are observed with atomically sharp silicon tips, where the current density is approximately $5.5\times 10^8$ A/m$^2$ \cite{Ding2002}, which is much lower than the value associated with the vacuum space charge saturation. 

Fig. \ref{fig:fe_result}(b) also demonstrates that the current saturation level is strongly dependent on the mobility of charge carriers, with higher mobilities resulting in higher saturation currents. This is a natural result of the SBLF-based model as saturation is linked to the efficiency of charge drift through the depleted region resulting in the relatively low current density of \ce{Cs2Te}.

\subsection{Heat Dissipation due to Field Emission}

The magnitude of Joule heating will depend on the the electrical resistivity of the material. Since the resistivity of \ce{Cs2Te} is not well known experimentally, it was calculated using the canonical relation
\begin{equation}\label{resist}
    \rho = \frac{1}{q(n\cdot\mu_n + p\cdot\mu_p)}.
\end{equation}
with electron and hole mobility values obtained from density functional theory (DFT) elsewhere \cite{Gaoxue2024}. Eq.~\ref{resist} suggests a range for resistivity between 0.1 and 30 $\Omega\cdot$m. In the present work, 30 $\Omega\cdot$m was utilized to simulate the worst-case scenario with the maximum amount of potential heating.

\begin{figure}
	\includegraphics[width=8.6cm]{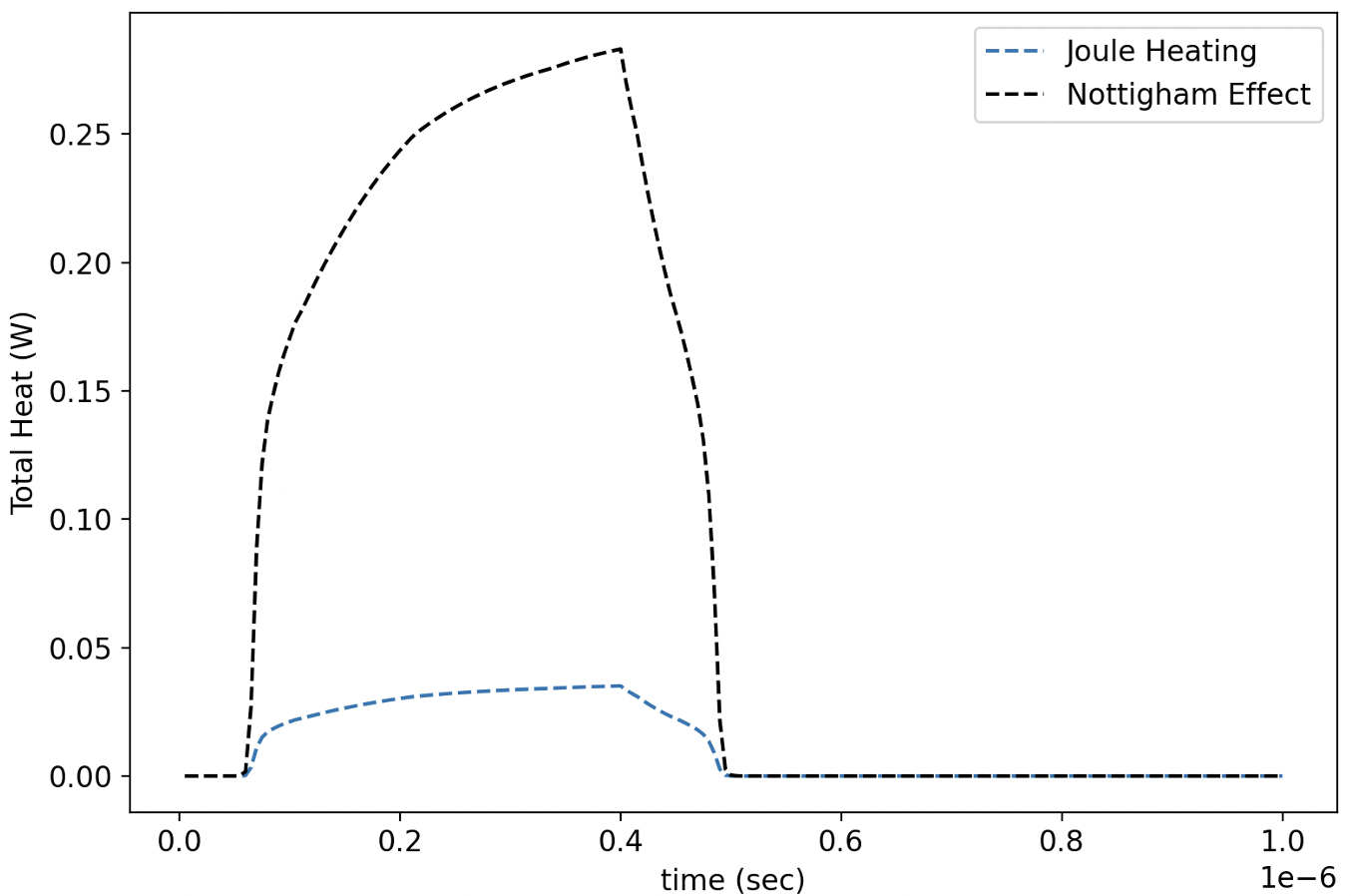}
	\caption{Total generated heat for Nottingham effect (black) and Joule effect (blue) during a single RF pulse.} 
	\label{fig:parameter_analysis}
\end{figure}

The individual transient contributions of Joule and Nottingham effects to \ce{Cs2Te} heating are displayed in Fig. \ref{fig:parameter_analysis}. 
The results show that the Nottingham effect dissipates about an order of magnitude more power than Joule heating. This is expected under low (due to charge depletion saturation) current density conditions \cite{Nottingham_CNT}, as Nottingham power scales as $j$ and Joule power scales as $j^2$.

\subsection{Heating Behavior: Single Pulse}
\begin{figure*}
	\includegraphics[width=\textwidth]{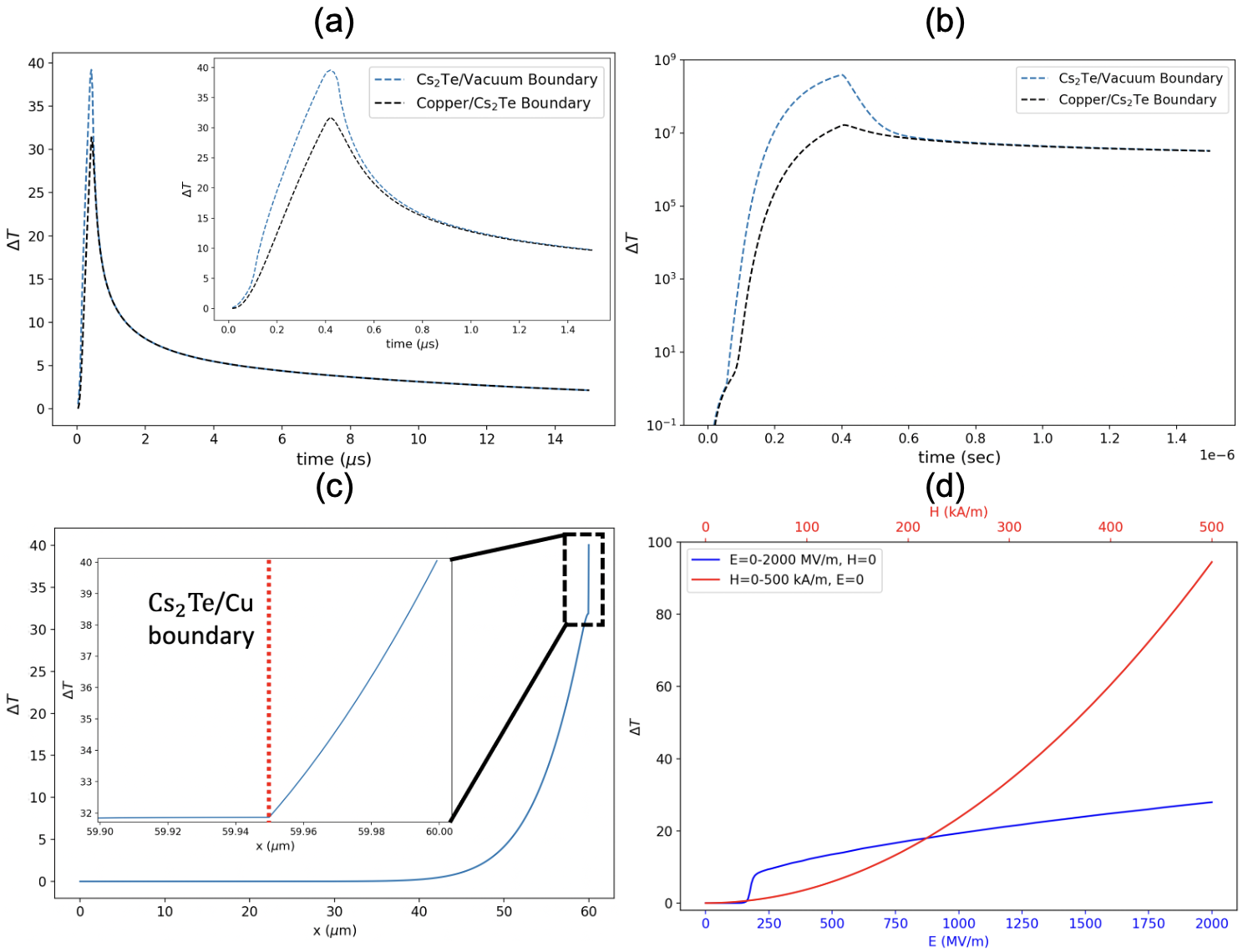}
	\caption{(a) Relative temperature rise under a single RF pulse. Here, current density follows the SBLF model. (b)  As in (a) but if the current density followed the FN model. (c) 1D spatial temperature distribution at $t$=0.4 $\mu$s. (d) Maximal relative temperature increase when independently varying applied magnetic field and electric field.}
	\label{fig:sim_result}
\end{figure*}


\begin{table}[!]
 \centering
\caption{Summary of Material Parameters}
\begin{tabular}{ l | l | l | l}

Parameter & Units & Copper & \ce{Cs2Te} \\
\hline
\hline
Heat Capacity & $C_e$ [$J$ $kg^{-1}$ K$^{-1}$]          & 375                  & 490        \\
Mass Density & $\rho_m$ [kg m$^{-3}$]       & 8960                 & 6150       \\
Thermal Conductivity & $\kappa$ [W m$^{-1}$ K$^{-1}$]      & 401                  & 0.12       \\
Skin Depth & $\delta$ [$\mu$m]     & 0.595         & 840 \\
Electrical Resistivity & $\rho_e$ [$\Omega\cdot$m] & 1.77e-8   & 30 \\
Electron Mobility & $\mu_e$ [cm$^2$V$^{-1}$s$^{-1}$] &  & 20 \\
Layer Thickness & $L$ [m] &  60 $\mu$m               & 50 nm      \\               

\end{tabular}
\label{table:material}
\end{table}

\begin{table}[!]
 \centering
\caption{Summary of Simulation Parameters}
\begin{tabular}{ l | l | l}

Parameter & Units & Value \\
\hline
\hline
Repetition Rate & $f$ [Hz]         & 375                  \\
RF frequency & $f_{rf}$ [GHz]   &  5.36 \\
Pulse Length & $t_p$ [ns]      & 400               \\
Surface E-field & $E_0$ [MV/m]    & 300                  \\
Surface H-field & $H_0$ [kA/m]    & 400  \\
Fill Time & $\tau$ [ns] & 100 \\
Transfer Coefficient & $h$ [Wm$^{-2}$K$^{-1}$]  & 1.2e4 \\

\end{tabular}
\label{table:sim_parameter}
\end{table}

The combined contribution of pulse heating and field emission (including both Joule and Nottingham contributions) on the temperature evolution in the material can now be investigated. The input parameters used for these simulations are summarized in Table \ref{table:material} and Table \ref{table:sim_parameter}.

Fig. \ref{fig:sim_result}(a) reports the temperature evolution due to a single RF pulse on the Cu/\ce{Cs2Te} boundary (black) and the \ce{Cs2Te}/vacuum boundary (blue). For a surface electric field of 300 MV/m and a magnetic field of 400 kA/m, the photocathode system reached a maximum $\Delta T$ of $\sim$40 K at the \ce{Cs2Te}/vacuum boundary and of $\sim$30 K at the Cu/\ce{Cs2Te} boundary. After the end of the pulse, the system rapidly ($\sim 10$ $\mu$s) cool down  to the initial temperature of the surrounding. 

In striking comparison, Fig. \ref{fig:sim_result}(b) reports the temperature rise when field emission  is instead described using FN theory. In this case, the system would undergo a catastrophic failure  due to explosive temperature rise. This is consistent with the scenario described by 
 Kyritsakis $et$ $al.$ \cite{kyritsakis2018thermal} for sharp Cu emittors. Capturing the current saturation behavior accurately is therefore critical to correctly estimate the heat dissipation in this system and hence to realistically predicts its stability under high RF power conditions. 

Fig.\ref{fig:sim_result}(c) reports the temperature distribution throughout the entire computational domain at $t$=400 ns when the pulse is turned off. The \ce{Cs2Te}/vacuum boundary is located at $x$=60 $\mu$m while the Cu/\ce{Cs2Te} boundary is located at $x$=59.95 $\mu$m. The differences in volumetric heat capacity and thermal conductivity between the two materials results in a kink in the temperature profile at the metal/semi-conductor interface. 
The results show a temperature increase is particularly significant in the topmost $\sim$10 $\mu$m closest to the vacuum interface. 

To gain a better understanding of the individual contributions to the overall heating, a set of simulations was conducted where the applied magnetic field was varied from 0 to 500 kA/m while keeping the electric field at 0 V/m, and another set of simulations where the applied electric field was varied from 0 to 2,000 MV/m while keeping the magnetic field at 0 A/m. 
The results are presented in Fig. \ref{fig:sim_result}(d). Both curves exhibit significantly different scaling behavior. The saturation of the electric-field driven field emission currents predicted by the SBLF theory lead to a peak temperature increases between 10 and 30 K between about 225 MV/m and 2 GV/m, and to very low heating below. In contrast, pulse heating driven by magnetic fields results in a quadratic increase in heat dissipation, leading to a crossover between and electric/field emission-dominated regime at low magnetic fields and a magnetic/pulse-heating-dominated regime at high magnetic fields. 

\subsection{Heating Behavior: Multiple Pulses}
The single-pulse results presented above can be extended to the (common) scenario where  multiple RF pulses (higher duty cycle operation) are applied in succession, again using the material and simulation parameters listed in in Table \ref{table:material} and Table \ref{table:sim_parameter}, with the exception of the Cu layer thickness being increased to 10 mm to capture a more realistic heat dissipation transients at the back-plate boundary, which is important at longer simulation times. This larger domain size, however, made explicit simulations computationally onerous. To make the calcualtion more affordable,  the heat diffusion equation was solved using time-independent heat sources corresponding to their temporally-averaged values across the RF pulse sequence, following Ref.\ \onlinecite{analyticalRFpulsedheating}.
These values were calculated numerically by integrating Eq. \ref{eq:P(x,t)} and multiplying it by the pulse frequency to obtain an average heat dissipation per unit time as

\begin{align}
    q_{0}(x)&=f \int_0^{t_{\text{end}}} P(x,t) dt \notag \\
         &=f \frac{R_s}{\delta}e^{-2x/\delta} ( \int_0^{t_p} H_0 (1-e^{-t/\tau}) dt   \notag \\
         &+ \int_{t_p}^{t_\text{end}} H_0 (1-e^{-t_p/\tau})e^{-(t-t_p)/\tau} dt)
\end{align}

\begin{figure}
	\includegraphics[width=8.6cm]{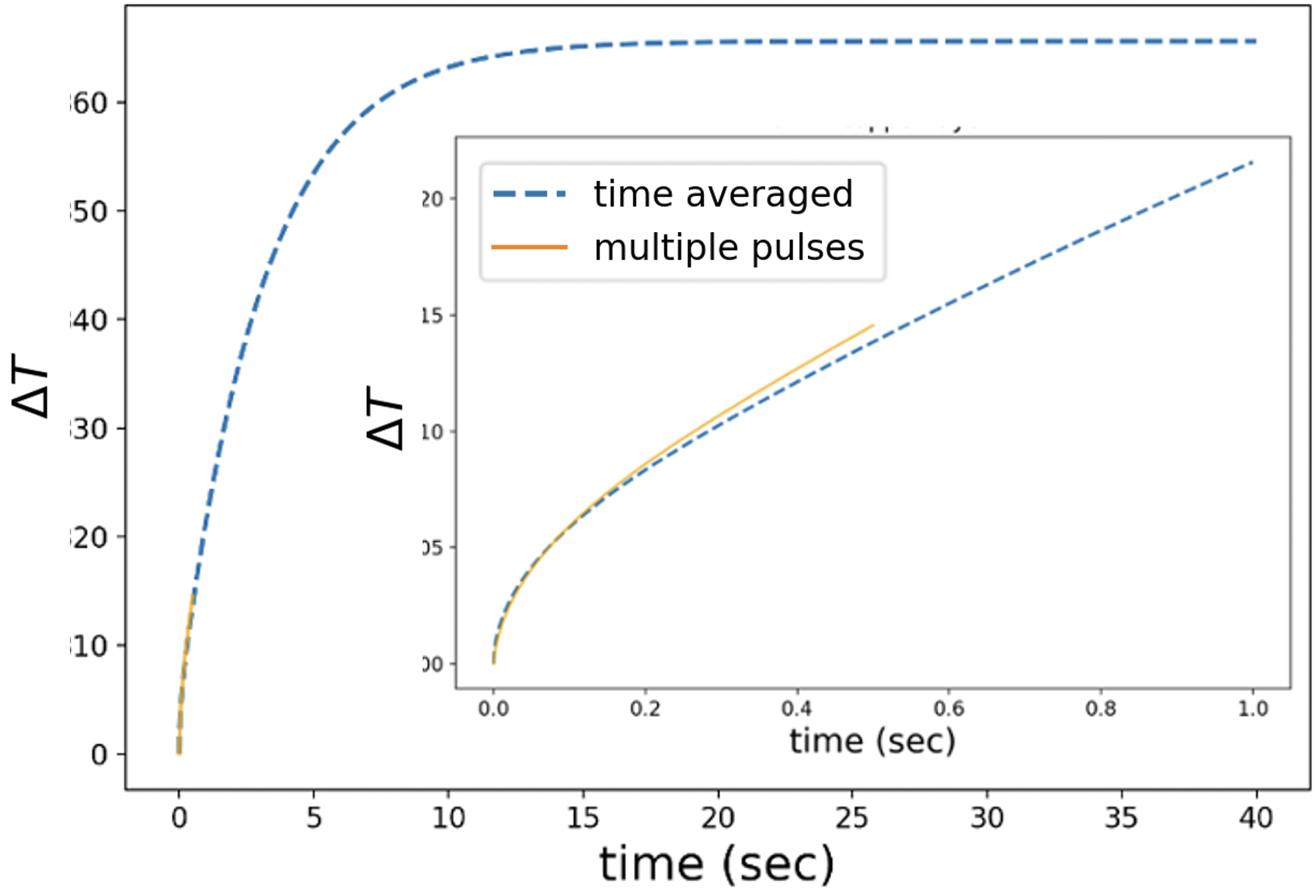}
	\caption{Maximal attainable temperature rise with explicit calculation of the temperature spike over multiple pulses (orange) vs calculation using time-averaged heat source (blue).}
	\label{fig:multiple_pulse}
\end{figure}

Fig. \ref{fig:multiple_pulse} shows the resulting temperature evolution at the \ce{Cs2Te}/vacuum boundary utilizing this time-averaged strategy, shown with blue dashed line, which can be compared to a shorter explicit calculation without intra-pulse averaging shown in orange., showing very good agreement. Extending the calculation until a steady state is reached leads to a predicted temperature increase of around 60K. 

\label{subsec::multi-pulse}

\subsection{Beyond common scenarios}
The simulations presented above suggested that RF pulsed heating combined with field emission heating would lead to a relatively limited temperature elevation on the semiconductor photocathode surface. Additionally, due to field emission current saturation, processes leading to runaway and catastrophic failure appear significantly less efficient in semi-conductors than in metals where saturation of the field emission currents is not expected. However, further analysis may be required to understand the emission characteristic for high-gradient application of semiconductor photocathodes where operational fields exceed 100 MV/m.

Field emission from semiconductor are indeed known to experience current saturation when the surface field is increased beyond a certain point, which limits heating contributions from field emission. However, the appearance of a different regime at ultrahigh gradients should also be considered.
In DC experiments, it was indeed shown that at progressively higher surface fields exceeding the saturation point, the $I-V$ can switch from saturation to a highly nonlinear regime \cite{Serbun_silicon2012} where the 
field emission current increases rapidly.
It was hypothesized that this rapid increase in electron emission results from thermionic emission \cite{Binh2001}, pointing toward the possibility of a prethermal runaway/explosive emission stage.
Another alternative explanation is that dielectric breakdown could lead to a similar increase in field emission when the local fields in the material become large enough to trigger avalanches of impact ionization, leading to the formation of a hot free carriers gas and to the restoration of an FN-like field emission behavior.

Further, we must relate the dielectric breakdown effect and high gradient application in RF environment. Unlike DC applications, the ramp rate of the voltage flowing through the film must be considered. Johnson's figure of merit (JFOM) characterizes the material's suitability for high power and high frequency application. 
JFOM gives the theoretical limits of a semiconductor's performance in terms of power handling and voltage ramp rate before dielectric breakdown occurs. JFOM is defined as
\begin{equation}
    \text{JFOM}=\frac{E_{br} \cdot v_{sat}}{2\pi},
\end{equation}
where $E_{br}$ is the semiconductor breakdown field and $v_{sat}$ is the saturation drift velocity. For \ce{Cs2Te}, $E_{br}=\sqrt{\frac{2\cdot q\cdot n\cdot E_g/4}{\varepsilon_0\varepsilon}}$ \cite{Sze2006} can be predicted as $\approx$10 MV/m at a carrier concentration of $10^{16}$ cm$^{-3}$ and 300 K. $v_{sat}$ of \ce{Cs2Te} is not well documented but this value is relatively consistent among various semiconductors and at around $10^5$ m/s. This results in a JFOM of \ce{Cs2Te} around $\sim$0.15 V/ps.
In C-band ($f_{RF}\approx 5$ GHz) the applied electric field ramps up to it's maximum in 50 ps (1/$f_{RF}$/4). Assuming a maximum applied field of 300 MV/m and a dielectric constant of 10, the ramp rate of the electric field gradient would becomes $1.5\times 10^6$ V m$^{-1}$ps$^{-1}$. With the typically used film thickness of 50 nm, the ramp rate becomes 0.075 V/ps for a \ce{Cs2Te} film photocathode. This is only a factor of 2 below the calculated JFOM limit, which suggests the possibility of \ce{Cs2Te} going into avalanche breakdown under the described operating conditions.
The voltage ramp rate approaching the JFOM limit opens up an interesting research avenue.  Avalanche forming in \ce{Cs2Te} would create a metallic-like electron plasma. This plasma could then be extracted by the trailing portion of the peaking electric field, thereby inducing heat release and temperature spike comparable in magnitude with the one shown in Fig.\ref{fig:sim_result}b. This could potentially then result in thermal runaway and in breakdown/arc formation.

\section{Conclusion}

This study computationally investigated the thermal balance in a thin \ce{Cs2Te} photocathode film on metal operating under very high fields in order to assess its susceptibility to conventional thermal runaway leading to an electrical breakdown in vacuum. 
It was found that under planned upgraded application conditions, beyond the state of the art 100-150 MV/m, thermal-runaway is impeded by field emission current saturation. At the same time, it was found that the averaged pulse heating is also unlikely to yield catastrophic material failure.

However, peak temperature spikes and steady-state temperature rise can be expected between 70 and 100 K. Such temperature elevation on the photocathode surface, when taken together with very high electric fields, were recently shown to generate a non-negligible thermo-elastic driving force that can result in 
diffusive roughening of the surface \cite{Shinohara2024}. Previous research indicates that such thermal-mechanical effect assists in the nucleation of the geometrical rf-breakdown precursors on metal surfaces and could potentially also be of importance to metal-coated semiconductor electron emission cathodes. Modeling of metal-semiconductor thermo-elastic in high gradients are underway.

Additionally, the present work highlights the risk of avalanche breakdown \ce{Cs2Te} that could potentially lead to metal-like thermal-runaway \cite{2018Kyritsakis}, subsequently triggering rf breakdown. Time-resolved multi-physics modeling of dielectric breakdown \cite{herrmann2023computationally}, that could potentially become a limiting factor in very high gradient semiconductor cathode system, is being developed.

\section*{Acknowledgments}
Ryo Shinohara, Soumendu Bagchi, Evgenya Simakov, and Danny Perez were supported by the Laboratory Directed Research and Development program of Los Alamos National Laboratory under project number 20230011DR.
The work by Sergey Baryshev was supported by the U.S. Department of Energy, Office of Science, Office of High Energy Physics under Award No. DE-SC0020429.
Los Alamos National Laboratory is operated by Triad National Security, LLC, for the National Nuclear Security Administration of U.S. Department of Energy (Contract No. 89233218CNA000001).
\bibliography{references}

\end{document}